\documentclass[conference]{IEEEtran}
\IEEEoverridecommandlockouts
\usepackage{stfloats}
\usepackage{tabularx}
\usepackage{cite}
\usepackage{amsmath,amssymb,amsfonts}
\usepackage{algorithmic}
\usepackage{graphicx}
\usepackage{textcomp}
\usepackage{xcolor}
\def\BibTeX{{\rm B\kern-.05em{\sc i\kern-.025em b}\kern-.08em
    T\kern-.1667em\lower.7ex\hbox{E}\kern-.125emX}}
\usepackage{booktabs}

\usepackage{booktabs} 
\usepackage{caption} 

\usepackage[normalem]{ulem}   
\usepackage{xcolor}           

\usepackage{cancel}           

\newif\ifshowchanges 
\showchangestrue   

\DeclareRobustCommand{\del}[1]{%
  \ifshowchanges
    \ifmmode \cancel{#1}\else \sout{#1}\fi
  \fi
}

\usepackage[most]{tcolorbox}

\usepackage[hyphens]{url}
\usepackage[hidelinks]{hyperref}
\usepackage{microtype}

\begin{document}

\title{From Reviewers’ Lens: Understanding Bug Bounty Report Invalid Reasons with LLMs}

\author{
\IEEEauthorblockN{Jiangrui Zheng, Yingming Zhou, Ali Abdullah Ahmad, Hanqing Yao, and Xueqing Liu\textsuperscript{\textdagger}}
\IEEEauthorblockA{
Department of Computer Science\\Stevens Institute of Technology\\Hoboken, NJ, USA\\
\{jzheng36, yzhou136, aahmad6, hyao9, xliu127\}@stevens.edu
}
\thanks{\textsuperscript{\textdagger}Corresponding author.}
}

\maketitle

\begin{abstract}

Bug bounty platforms (e.g., HackerOne, BugCrowd) leverage crowd-sourced vulnerability discovery to improve continuous coverage, reduce the cost of discovery, and serve as an integral complement to internal red teams. With the rise of AI-generated bug reports, little work exists to help bug hunters understand why these reports are labeled as invalid. To improve report quality and reduce reviewers' burden, it is critical to predict invalid reports and interpret invalid reasons.

In this work, we conduct an empirical study with the purpose of helping bug hunters understand the validity of reports. We collect a dataset of 9,942 disclosed bug bounty reports, including 1,400 invalid reports, and evaluate whether state-of-the-art large language models can identify invalid reports. While models such as GPT-5, DeepSeek, and a fine-tuned RoBERTa achieve strong overall accuracy, they consistently struggle to detect invalid cases, showing a tendency to over-accept reports.
To improve invalidity detection, we build a taxonomy of rejection reasons for Information Disclosure vulnerabilities and incorporate it into a retrieval-augmented generation (RAG) framework. This approach substantially improves classification consistency and reduces bias. We also examine whether reviewer decisions may be influenced by factors beyond the content of the report. Our analysis shows that reporters with higher reputations tend to receive more favorable outcomes in borderline cases, suggesting that perceived expertise can influence review judgments.

Overall, our findings highlight the challenges of invalid report identification and show that combining LLMs with structured reviewer knowledge can support more transparent and consistent vulnerability report review.

\end{abstract}


\section{Introduction}


\begin{figure}[t]
\centering
\includegraphics[width=0.48\textwidth]{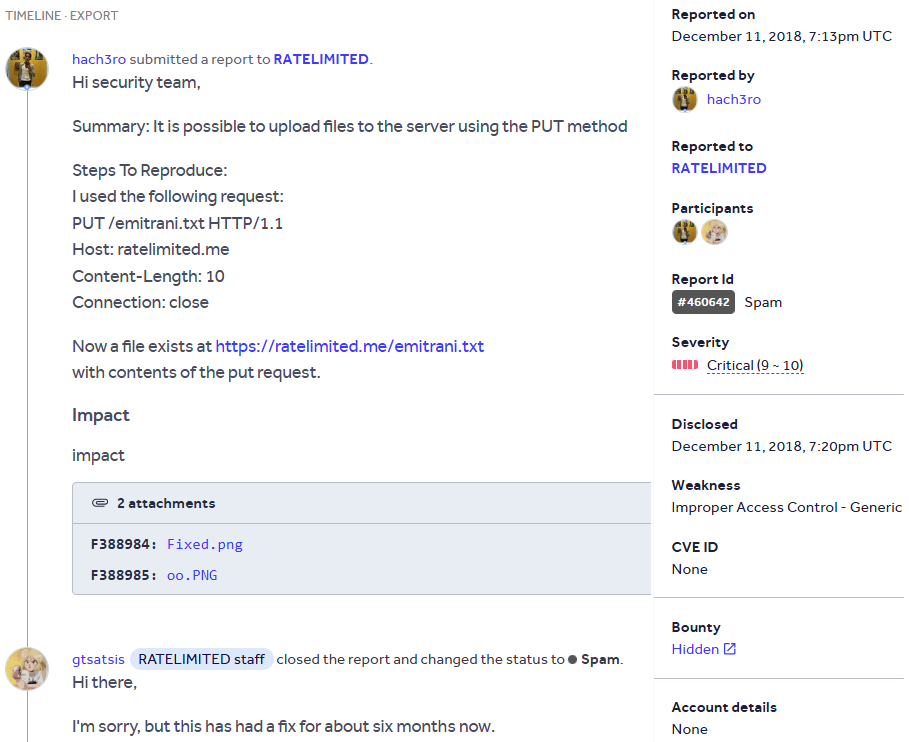} 
\vspace{-0mm}
\caption{Example invalid bug bounty report (\#460642, HackerOne~\cite{hackerone_460642}). 
The submission, titled ``HTTP PUT method enabled,'' was closed as \textit{Spam} because it reused a previously disclosed demo.}
\label{fig:invalid_example}
\vspace{-3mm}
\end{figure}

“Given enough eyeballs, all bugs are shallow.” This well-known principle, often referred to as Linus’ Law, has been empirically supported by analyses of bug bounty programs \cite{maillart2017given}. 
Bug bounty programs (e.g., HackerOne, BugCrowd) leverage crowd-sourced vulnerability discovery to improve continuous coverage, reduce the cost of discovery, and serve as an integral complement to internal red teams. 
However, as vulnerability discovery activities scale up, there is a growing expectation gap between bug hunters and reviewers (i.e., vendor staff). 
Prior studies have shown that reviewing bug bounty submissions is highly resource-intensive and time-consuming for organizations~\cite{walshe2022coordinated}. There is an increasing number of spam or AI-generated reports that pollute the overall bug bounty ecosystem and make it harder to identify genuine vulnerabilities. From the reviewers' perspective, these programs are increasingly challenged by a surge of invalid or low-quality submissions. Curl project founder Daniel Stenberg, for instance, publicly expressed frustration at the overwhelming number of AI-generated bug reports that waste valuable reviewing efforts and slow down the overall vulnerability handling process\cite{theRegister2025CurlAIBugReports}.


To the best of our knowledge, limited studies have investigated how to help hunters predict and understand the invalidity of their reports before submission. The closest work is by Shafigh et al.~\cite{shafigh2021some}, but they focus only on out-of-scope invalid reports, without addressing other types of invalidity, and they do not analyze valid submissions. Moreover, few studies have examined whether modern large language models (LLMs) can effectively support reviewers in making fair and consistent decisions.
With their strong capacity for textual reasoning and contextual understanding, LLMs present a promising avenue for modeling how reviewers assess report validity.


To shed light on the invalidity reasons, we conduct an empirical study using all the publicly disclosed vulnerability reports from HackerOne, one of the leading security bug bounty platforms, where vendors pay cash to reward third-party security researchers' efforts in helping them find bugs. The security bugs covered include web, mobile, cloud, IoT, etc. We collect 9,942 vulnerability reports from HackerOne (8,542 valid and 1,400 invalid). 

We explore the following research questions:




\noindent \textbf{RQ1: How effective are LLMs in predicting invalidity?}\\
We first benchmark state-of-the-art LLMs, including GPT-5 and DeepSeek, alongside a fine-tuned RoBERTa model to evaluate their baseline ability to classify bug bounty reports as valid or invalid.

\noindent \textbf{RQ2: Does incorporating reviewer knowledge help predict invalidity?}\\
Next, we investigate incorporating three aspects of reviewer knowledge: vendor scope definitions, program policies, and a taxonomy of invalid report types. We integrate the taxonomy using a retrieval-augmented generation (RAG) framework to models.

\noindent \textbf{RQ3: Does bias exist on reviewers' invalidity judgment?} \\
Finally, we conduct a fairness analysis to examine whether reviewer decisions are influenced by non-content factors such as reporter reputation.
We identify pairs of reports within the same weakness category that are semantically similar but have different validation statuses.
By comparing these pairs, we analyze the extent to which reputation differences correlate with outcome discrepancies.

Our findings for \textbf{RQ1}--\textbf{RQ3} are summarized as follows:
\begin{tcolorbox}[
    colback=white,        
    colframe=black,       
    arc=5pt,              
    boxrule=0.7pt,        
    left=4pt, right=10pt,  
    top=6pt, bottom=6pt   
]
\begin{enumerate}
    \item LLMs such as GPT-5, DeepSeek, and fine-tuned RoBERTa model show strong overall accuracy but struggle to identify invalid reports, revealing a bias toward acceptance.

    \item Adding reviewer knowledge in the form of a taxonomy of invalid report types helps the models detect invalid cases more reliably and make decisions more consistently.

    \item Review outcomes are not determined solely by the content itself. High-reputation reporters receive more favorable outcomes even for reports that are nearly identical to those from low-reputation reporters.
\end{enumerate}
\end{tcolorbox}

Together, these results provide new insights into the human factors and reasoning cues behind bug bounty validity decisions.


\section{Background and Motivation }

\subsection{How Bug Bounty Programs Work}
\noindent\textbf{Overview of Bug Bounty Platforms} Bug bounty platforms are gaining widespread attention across industry, government, and academia\cite{bohme2006comparison,laszka2016banishing,ozment2004bug}. A bug bounty program is a structured vulnerability disclosure mechanism that allows independent security researchers to report security flaws to organizations in exchange for monetary rewards or recognition. As described by HackerOne~\cite{hackerone_2024_bugbounties}, companies define eligible assets and vulnerability types (\emph{scope}), and offer bounties based on the severity and impact of verified findings. 

\noindent\textbf{Workflow of Bug Bounty Process.} 
The process typically involves four stages: (1) researchers discover and submit vulnerabilities, (2) reviewers validate the reports and determine their legitimacy, (3) vendors fix confirmed issues, and (4) rewards are issued according to program policy. Reports that are out of scope, invalid, or have negligible impact are rejected without reward. 

This workflow enables organizations to continuously test and improve their security through crowdsourced expertise while relying on consistent review and reward mechanisms managed by the platform.

\noindent\textbf{Major Bug Bounty Platforms.}
Bug bounty platforms such as HackerOne and Bugcrowd mainly host vendor-specific programs, where companies define their own scopes and review reports internally. 
In contrast, Huntr\cite{huntr} focuses on vulnerabilities in open-source projects: researchers still submit reports through the platform, but the issues are routed to project maintainers (e.g., via GitHub). Prior studies have investigated multiple dimensions of bug bounty platforms.


\subsection{Challenges in AI-driven Bug Hunting Era}

Recent years have brought a surge of AI-generated submissions in bug bounty programs.
Many of these reports appear convincing at first glance but contain fabricated evidence or shallow reasoning, creating significant overhead for triage teams.
High-profile maintainers describe being “effectively DDoS’ed” by low-effort, machine-written vulnerability claims, prompting them to adopt pre-screening mechanisms and stricter submission filters to manage the flood of incoming reports~\cite{theRegister2025CurlAIBugReports}.

Multiple public accounts have documented large waves of such “AI-slop” reports across both open-source projects and commercial bounty platforms, where maintainers spend hours disproving technically invalid claims that nonetheless sound credible~\cite{newstack2025curlAIspam,ars2025curlAISlop,stenberg2025deathbyslop}.
At the same time, AI tools are increasingly used by hunters to explore new attack surfaces and automate vulnerability discovery~\cite{cso2025AIbughunting,hackerone2025BeyondTheNoise}.
While this accelerates legitimate findings, it also increases the total volume of submissions that reviewers must process~\cite{81m_hackerone2025}.
As language models continue to improve, surface-level cues will no longer suffice to identify unreliable reports, forcing reviewers to rely more on evidence verification—such as reproducing proof-of-concept code or confirming exploit chains—and highlighting the need for semi-automated triage support~\cite{hackerone2025BeyondTheNoise}.


Beyond AI-generated noise, reviewers continue to face familiar challenges that have persisted for years.
Differences in security expertise among maintainers often lead to inconsistent judgments on subtle exploitation details~\cite{hendrickmaintainer}.
Meanwhile, platforms still receive large numbers of duplicate or near-duplicate reports describing the same underlying issue, adding to the workload and slowing response times~\cite{zhao2016crowdsourced,ellis2022bounty,laszka2016banishing}.
While such challenges are not new, the growing number of submissions aided by AI tools has made them more prominent and harder to handle.

\noindent\textbf{Implications.}
The largely manual nature of bug bounty review, combined with the rise of low-quality and AI-generated submissions, highlights the need for more consistent and explainable validity judgments. Motivated by this need, our study investigates whether LLMs—when guided by structured reviewer knowledge—can help improve the reliability of the review process.

\section{Dataset}

\subsection{Data Collection}
We collect a total of 9,942 disclosed vulnerability reports from the HackerOne platform. Among these reports, 8,542 were accepted and 1,400 were not accepted. The dataset is randomly divided into training and test subsets at a 4:1 ratio while preserving the original distribution of valid and invalid reports, resulting in 7,953 training cases and 1,989 test cases.
Each report contains multiple metadata fields describing its submission details, review history, and reporter profile.  
Figure~\ref{fig:invalid_example} shows an example report entry.  
To facilitate analysis, we extract the following fields:

\begin{itemize}
    \item \textbf{Basic Information:} Includes the report title, submission date, reporter name, target program (\texttt{Reported to}), and final disclosure date.
    
    \item \textbf{Vulnerability Attributes:} Includes the reported \texttt{Weakness} category, \texttt{CVE ID}, and \texttt{Severity}.
    
    \item \textbf{Review Timeline:} Records all reviewer–hunter interactions, including comments, status changes, and disclosure events.
    
    \item \textbf{Reporter Profile:} Extracted from the reporter’s public profile section, including long-term performance metrics such as \texttt{Signal}, \texttt{Impact}, and \texttt{Reputation}.

    \item \textbf{Final Status:} Indicates the triage outcome assigned by reviewers 
    (\textit{Resolved}, \textit{Informative}, \textit{Not Applicable}, \textit{Spam}, etc.).  
    Reports marked as \textit{Resolved} are treated as valid and fixed vulnerabilities, 
    while those labeled \textit{Not Applicable} or \textit{Spam} are considered invalid.  
    The \textit{Informative} status refers to submissions that contain useful information but do not require any security action~\cite{hackerone_report_states}.
\end{itemize}


\noindent \subsection{Data Distribution Analysis}

\noindent \textbf{Vulnerability Types.} 
Table~\ref{tab:vuln-types} lists the top ten reported weaknesses. 
While the table presents disaggregated data, we observe that multiple variants of Cross-Site Scripting (XSS) together account for 13.8\% of all submissions. 
Information Disclosure remains the single most frequent weakness (9.4\%), followed by Access Control and Authentication issues, which together represent more than 10\% of the dataset.



\begin{table}[t]
\caption{Top 10 Vulnerability Types in HackerOne Reports}
\label{tab:vuln-types}
\centering
\small
\renewcommand{\arraystretch}{1.15}
\begin{tabular}{|c|p{5cm}|c|c|}
\hline
\textbf{Rank} & \textbf{Weakness Type} & \textbf{Count} & \textbf{\%} \\

\hline
1 & Information Disclosure & 931 & 9.4 \\
2 & Improper Access Control - Generic & 612 & 6.2 \\
3 & Cross-site Scripting (XSS) - Reflected & 483 & 4.9 \\
4 & Violation of Secure Design Principles & 461 & 4.6 \\
5 & Cross-site Scripting (XSS) - Stored & 451 & 4.5 \\
6 & Improper Authentication - Generic & 442 & 4.4 \\
7 & Cross-site Scripting (XSS) - Generic & 436 & 4.4 \\
8 & Uncontrolled Resource Consumption & 414 & 4.2 \\
9 & Business Logic Errors & 306 & 3.1 \\
10 & Cross-Site Request Forgery (CSRF) & 304 & 3.1 \\
\hline
\end{tabular}
\end{table}





\noindent \textbf{Target Organizations.} Table~\ref{tab:targets} lists the top programs. Government and defense (U.S. DoD 8.3\%) and open-source initiatives (Internet Bug Bounty 6.8\%) are heavily targeted.

\begin{table}[htbp]
\caption{Top 10 Target Programs In HackerOne report}
\label{tab:targets}
\centering
\small
\renewcommand{\arraystretch}{1.15}
\begin{tabular}{|c|p{4.3cm}|c|c|}
\hline
\textbf{Rank} & \textbf{Organization} & \textbf{Reports} & \textbf{\%} \\
\hline
1 & U.S. Dept of Defense & 828 & 8.3 \\
2 & Internet Bug Bounty & 677 & 6.8 \\
3 & Nextcloud & 504 & 5.1 \\
4 & HackerOne & 384 & 3.9 \\
5 & Shopify & 339 & 3.4 \\
6 & Node.js Modules & 303 & 3.0 \\
7 & GitLab & 250 & 2.5 \\
8 & curl & 180 & 1.8 \\
9 & X / xAI & 163 & 1.6 \\
10 & VK.com & 162 & 1.6 \\
\hline
\end{tabular}
\end{table}

\section{RQ1: How effective are LLMs in predicting invalidity?}

\subsection{Motivation} 
Although LLMs such as GPT-5 and DeepSeek have demonstrated strong reasoning ability across many domains, it remains unclear whether they can generalize to the nuanced review process of bug bounty platforms, where rejection decisions often depend on implicit reviewer knowledge.  
Understanding their baseline capability is crucial before introducing any reviewer-specific context or knowledge augmentation.

\subsection{Methodology}

We first examine the capability of state-of-the-art large language models to review bug bounty reports. Specifically, we evaluate GPT-5 and DeepSeek on the task of classifying HackerOne reports as valid or invalid. Each model is prompted to act as an expert bug bounty reviewer. The LLMs are zero-shot prompted using the following instruction:
\begin{tcolorbox}
You are an expert bug bounty reviewer.

Given the following HackerOne report, classify it strictly as: \\
1 = valid (valid vulnerability report), 0 = invalid (Informative, Not Applicable, Spam).

Report to evaluate:

$[report\_text]$

Output only:\\ 1 = VALID, 0 = INVALID

\end{tcolorbox}

To establish a supervised baseline, we fine-tune a RoBERTa model on the collected HackerOne reports. The model is trained with a learning rate of $2\times10^{-5}$, weight decay of 0.01, and 3 epochs.

\subsection{Experimental Results and Analysis}

\noindent\textbf{Metric Definition.}  
We treat \textbf{valid reports} as the positive class (\(P\)) and \textbf{invalid reports} as the negative class (\(N\)).  
Accordingly:  
\[
\begin{aligned}
\text{TP} &: \text{ valid reports correctly predicted as valid},\\
\text{FP} &: \text{ invalid reports incorrectly predicted as valid},\\
\text{FN} &: \text{ valid reports incorrectly predicted as invalid},\\
\text{TN} &: \text{ invalid reports correctly predicted as invalid}.
\end{aligned}
\]
\begin{itemize}
    \item \textbf{Accuracy (Acc):} $(\mathrm{TP} + \mathrm{TN}) / (\mathrm{TP} + \mathrm{FP} + \mathrm{FN} + \mathrm{TN})$.
    \item \textbf{Recall$_{v}$ / Recall$_{inv}$:}  
    $\mathrm{TP} / (\mathrm{TP} + \mathrm{FN})$ for valid class,  
    $\mathrm{TN} / (\mathrm{TN} + \mathrm{FP})$ for invalid class.
    \item \textbf{Precision$_{v}$ / Precision$_{inv}$:}  
    $\mathrm{TP} / (\mathrm{TP} + \mathrm{FP})$ for valid class,  
    $\mathrm{TN} / (\mathrm{TN} + \mathrm{FN})$ for invalid class.
    \item \textbf{F1$_{v}$ / F1$_{inv}$:}  
    Harmonic mean of precision and recall for each class.  
    $F1 = 2 \times (\text{Precision} \times \text{Recall}) / (\text{Precision} + \text{Recall})$.
    \item \textbf{Macro-F1:} Average of F1$_{v}$ and F1$_{inv}$, representing overall balance between the two classes.
\end{itemize}
\vspace{5mm}

\noindent\textbf{GPT-5 and Deepseek.} As shown in Table \ref{tab:LLMs}, both large language models achieve high overall accuracy (GPT-5: 0.82; DeepSeek: 0.79), yet exhibit a strong bias toward predicting valid reports.
GPT-5 reaches a high recall on valid cases (0.91) but extremely low recall on invalid ones (0.28), indicating a false negative rate of roughly 72\%, i.e., most invalid reports are mistakenly accepted.
DeepSeek alleviates this bias slightly (invalid recall = 0.44) while maintaining comparable precision, suggesting improved discrimination between valid and invalid classes.
However, both models’ F1 scores for invalid reports (0.30–0.37) remain far lower than for valid ones (0.87–0.90), indicating that overall accuracy can mask substantial disparities in per-class performance.

\noindent\textbf{Fine-tuned RoBERTa.} 
Because LLMs show limited performance, we are interested in examining whether a fine-tuned RoBERTa can perform better. The fine-tuned RoBERTa baseline achieves a better balanced performance (Macro-F1 = 0.72). Although its recall on invalid reports (0.43) is comparable to DeepSeek’s, it retains both strong recall (0.96) and precision (0.93) for valid cases, resulting in higher reliability overall.  
RoBERTa also improves F1 on invalid reports to 0.50, suggesting that task-specific fine-tuning enables better recognition of linguistic and structural patterns characteristic of invalid submissions.
Nevertheless, the model still fails to correctly handle borderline or unclear reports, showing that imbalanced data and implicit reviewer cues remain major challenges for automated review.

Most misclassifications come from two situations: 
(1) out-of-scope submissions that look plausible as a vulnerability but lack scope knowledge, and 
(2) reports that lack sufficient textual detail. 
Some reports rely mainly on screenshots or video attachments, which provide too little textual content for an LLM to reason about. 
For example, Report \#226514~\cite{hackerone_226514} contains only a short description and a PNG image showing a full-path disclosure on \texttt{airship.paragonie.com}. 
Although the issue is real, the reviewer marked it as “Informative’’ because the affected asset was out of scope.  
Such cases motivate our use of vendor-scope information and retrieved reference examples in RQ2, which help the model better handle borderline cases that lack enough text or depend on contextual program rules.

\subsection{Summary of Findings}
Our results show that large language models can handle bug bounty report validation fairly well in general,
but they still struggle to identify invalid cases and often accept reports that only look like real vulnerabilities.
Fine-tuning a task-specific model such as RoBERTa helps reduce this bias and makes the predictions more balanced between valid and invalid reports. 
However, the persistent performance gap across classes suggests that the models rely mainly on surface-level linguistic patterns rather than deeper reviewer reasoning.


\begin{table*}[t]
\centering
\setlength{\tabcolsep}{4pt}
\renewcommand{\arraystretch}{1.15}
\caption{Models' performance on the test set and entire dataset (Valid treated as positive, Invalid as negative). }
\label{tab:LLMs}
\resizebox{0.8\textwidth}{!}{%
\begin{tabular}{lcccccccc}
\toprule
\textbf{Model} & \textbf{Acc} & \textbf{Rec$_{val}$} & \textbf{Rec$_{inval}$} & \textbf{Pre$_{val}$} & \textbf{Pre$_{inval}$} & \textbf{F1$_{val}$} & \textbf{F1$_{inval}$} & \textbf{Mac-F1} \\
\midrule
GPT-5 & 0.822 & 0.912 & 0.275& 0.885 & 0.338 & 0.898 & 0.303 & 0.601 \\
DeepSeek & 0.789 & 0.846 & \textbf{0.439} & 0.902 & 0.319 & 0.873 & 0.369 & 0.621 \\
RoBERTa (finetuned) & \textbf{0.881} & \textbf{0.956} & 0.425 & \textbf{0.910} & \textbf{0.610} & \textbf{0.932} & \textbf{0.501} & \textbf{0.717} \\
\midrule
GPT-5 (All) & 0.818 & 0.914 & 0.324 & 0.882 & 0.355 & 0.897 & 0.339 & 0.618 \\
DeepSeek (All) & 0.795 & 0.853 & 0.465 & 0.899 & 0.352 & 0.875 & 0.403 & 0.639 \\
\bottomrule
\end{tabular}
}
\end{table*}


\section{RQ2: Does incorporating reviewer knowledge help predict invalidity?}

\begin{table*}[!t]
\centering
\small
\caption{Taxonomy of rejection reasons for invalid HackerOne reports.}
\label{tab:taxonomy}
\setlength{\tabcolsep}{5pt} 
\renewcommand{\arraystretch}{1.15} 
\begin{tabularx}{\textwidth}{p{2.2cm}p{4.5cm}X}
\toprule
\textbf{Status} & \textbf{Category} & \textbf{Description} \\
\midrule
\textbf{Informative} 
& Technical Issues & Known bugs, false positives, or unreproducible problems that do not constitute security vulnerabilities. \\
& Third-party Related Issues & Issues caused by external dependencies or third-party services outside program control. \\
& Scope/Policy Issues & Reports outside the defined program scope or violating disclosure policies. \\
& Risk Assessment & Valid findings assessed as low impact or below the action threshold. \\
& Special Scenarios & Non-standard cases such as debug modes, joke submissions, or physical access requirements. \\
& Path Disclosure Related & Exposure of file paths or directories without security impact. \\
& Version Disclosure Related & Disclosure of version or build information that is public or non-sensitive. \\
& User/Data Exposure & Non-sensitive user or general data exposed intentionally or without risk. \\
& File/Directory Related & Public or properly access-controlled files misreported as sensitive. \\
& Configuration/Credentials Related & Test credentials or placeholder configurations without real security implications. \\
& System Function / Security Trade-off Design & Intended behavior or accepted design trade-offs balancing usability and security. \\
& Tool/Protocol Behavior & Expected behavior of underlying tools, frameworks, or infrastructure. \\
\midrule
\textbf{Not Applicable}
& Not Applicable & The report doesn't contain a valid reproducible issue, and the security implications have not been demonstrated. \\
\midrule
\textbf{Duplicate}
& Duplicate & Issues already reported or resolved under previous submissions. \\
\midrule
\textbf{Spam}
& Spam / Irrelevant & Reports without legitimate security relevance, incomprehensible content, or promotional material. \\
\bottomrule
\end{tabularx}
\end{table*}

\subsection{Motivation} 
The baseline model from RQ1 tends to label most reports as valid, showing a bias toward acceptance. We hypothesize that this bias stems from the model's lack of reviewer-level contextual knowledge, such as what kinds of reports are typically rejected and why. 
To address this gap, we investigate whether providing explicit reviewer knowledge can help LLMs make more consistent and accurate decisions when identifying invalid reports.

\subsection{Methodology}

We enrich the model input with two forms of structured context: (1) vendor scope, which specifies the coverage of assets that are considered valid for submission in each program, and (2) a taxonomy of invalid report types. 
As a preliminary study, we focus on the \emph{Information Disclosure} category (Table~\ref{tab:taxonomy}). We choose  \textit{Information Disclosure} because it appears most frequently in our dataset (9.4\%) and covers diverse report forms. Its prevalence and variability make it a suitable category for an initial analysis of reviewer consistency. 
The underlying methodology is not specific to this category and can be transferred to other vulnerability classes, particularly those where reviewers rely heavily on written explanations rather than executable proofs of concept.

\begin{figure}[h]
\centering
\includegraphics[width=0.50\textwidth]{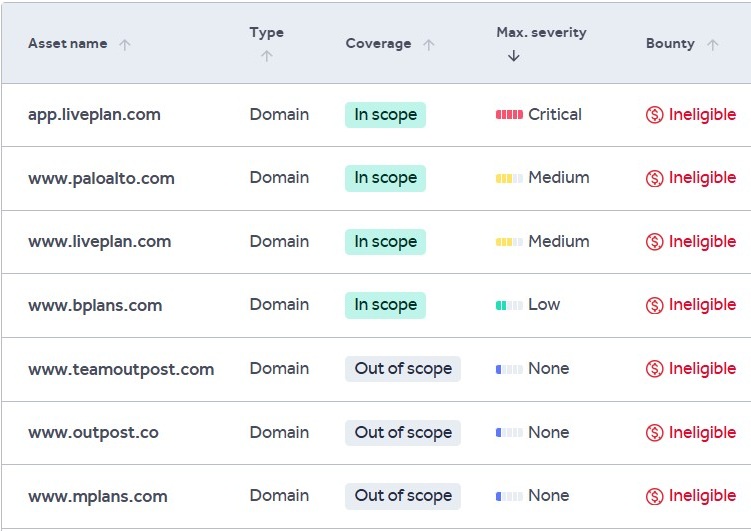} 
\vspace{-3mm}
\caption{Example of vendor scope entries on HackerOne (Palo Alto Software program)~\cite{hackerone_paloalto_policy_scopes}. Each entry defines the in-scope and out-of-scope assets for valid submissions.}
\label{fig:scope_example}
\end{figure}

\noindent\textbf{Vendor Scope Extraction.}
During inference, we use the vendor name in the report’s “Reported to” field to fetch the corresponding policy scope page and extract the raw scope entries. Figure~\ref{fig:scope_example} shows an example of the vendor scope table for the Palo Alto Software program on HackerOne.  
Each entry lists an \textit{asset name}, its \textit{type}, and the \textit{coverage status} indicating whether the asset is \textit{In Scope} or \textit{Out of Scope}.  
Additional metadata, such as severity limits and bounty eligibility, are available but not used in our model. We focus instead on coverage information, which defines valid submission boundaries.

To incorporate vendor policies into the model input, we insert the following vendor-scope instructions into the \texttt{$[vendor\_scope]$} placeholder of the main evaluation prompt:

\begin{tcolorbox}
\small
\textbf{Legend:}
\begin{itemize}
\item \textbf{Asset name:} Asset or endpoint name.
\item \textbf{Type:} Asset type such as \texttt{Domain}, \texttt{Source code}, or \texttt{Wildcard}.
\item \textbf{Coverage:} Whether this asset is in program scope (\texttt{True = In Scope}, \texttt{False = Out of Scope}).

\end{itemize}
- Asset name: [asset name 1] \textbar\ Type: [asset type 1] \textbar\ Coverage: [True or False]


% ...

\end{tcolorbox}

\noindent\textbf{Taxonomy RAG.}
To enhance the reasoning of the model with the knowledge of the reviewer, we design a retrieval-augmented generation (RAG) framework that integrates real examples of previously reviewed reports.  
The intuition is that by exposing the LLM to concrete invalid and valid cases with known outcomes, it can reason by analogy rather than relying purely on superficial text features, thereby reducing hallucination.

We first build a reference knowledge base from manually annotated \emph{Information Disclosure} reports. Each taxonomy type includes:  
(1) A short textual description of the invalid reasons, and  
(2) Several representative invalid reports that exemplify this reason.  

For each test report belonging to the \emph{Information Disclosure} category, we retrieve the top-$k$ most semantically similar reports ($k=3$) from the knowledge base using cosine similarity over text embeddings. 
We embed each report using the \texttt{sentence-transformers/all-mpnet-base-v2} model from the SentenceTransformer library and use normalized embeddings to compute similarity. 
We adopt a similarity threshold of $0.8$. This value falls within the range commonly used to identify near-duplicate or paraphrastic sentences in embedding-based similarity work, where higher cosine scores are shown to reflect strong semantic closeness~\cite{reimers2019sentence}.
To check the validity of this threshold, we manually review a sample of similarity-based pairs within the \textit{Information Disclosure} category and confirm that it captures reports describing essentially the same issue.
To maintain contextual consistency, retrieval is restricted to reports under the same weakness label, excluding unrelated or duplicate entries. 
For each retrieved report, we include its title and full content as contextual evidence. If the retrieved report is invalid, we additionally add its taxonomy type and corresponding description. This structured retrieval allows LLMs to recognize patterns of invalid reasoning and to ground its decision on actual reviewer judgments rather than surface-level keywords. During inference, GPT-5 receives the vendor’s in-scope information (if available), the retrieved reference reports, and the new report to be evaluated.
The model is then instructed to classify the report as either \textbf{VALID (1)} or \textbf{INVALID (0)} based on all provided information. 

The main evaluation prompt we use follows the instructions below:

\begin{tcolorbox}
You are an expert bug bounty assistant.

Vendor's in-scope assets and policies:
$[vendor\_scope]$

Below are example reports retrieved from the knowledge base that are semantically similar to the current report. Each example includes the report content, type, and description.

Use them as reasoning references — do not classify them again.

Retrieved reference reports:\\
$[rag\_reports\ (+type,\ +desc)\ if\ invalid]$

Now, determine whether the following new report is VALID (1) or INVALID (0).

Report to evaluate:
$[report\_text]$

Output only: \\
1 = VALID, 0 = INVALID

\end{tcolorbox}

\begin{table*}[t]
\centering
\setlength{\tabcolsep}{4pt}
\renewcommand{\arraystretch}{1.15}

\caption{Performance comparison on the \emph{Information Disclosure} subset. 
Valid is treated as positive and Invalid as negative.}
\label{tab:ID_rq2}
\resizebox{0.8\textwidth}{!}{%
\begin{tabular}{lcccccccc}
\toprule
\textbf{Setting} & \textbf{Acc} & \textbf{Rec$_{val}$} & \textbf{Rec$_{inval}$} & \textbf{Pre$_{val}$} & \textbf{Pre$_{inval}$} & \textbf{F1$_{val}$} & \textbf{F1$_{inval}$} & \textbf{Mac-F1} \\
\midrule
GPT-5 & 0.781 & 0.872 & 0.293 & 0.868 & 0.299 & 0.870 & 0.296 & 0.583 \\
{+}Scope & 0.699 & 0.768 & 0.333 & 0.861 & 0.211 & 0.812 & 0.259 & 0.535 \\
{+}Tax-RAG & \textbf{0.885} & \textbf{0.891} & 0.856 & \textbf{0.971} & \textbf{0.592} & \textbf{0.929} & \textbf{0.700} & \textbf{0.815} \\
{+}Tax-RAG+Scope & 0.803 & 0.791 & \textbf{0.863} & 0.969 & 0.434 & 0.871 & 0.578 & 0.725 \\
\bottomrule
\end{tabular}
}
\end{table*}

\subsection{Experimental Results and Analysis}
We compare four settings on the same subset: (i) the GPT-5 baseline (report only), (ii) \textbf{+Scope} (vendor scope added), (iii) \textbf{+Taxonomy RAG}, and (iv) \textbf{+Taxonomy RAG + Scope}. 
The baseline achieves high accuracy on valid cases but low recall on invalid ones, confirming a strong acceptance bias.

Adding \textbf{vendor scope} alone provides limited improvement and, in some cases, slightly reduces accuracy. This is likely because the vendor scope descriptions are often generic or high-level (e.g., program URLs or IP ranges) and provide little semantic guidance for distinguishing invalid content.

In contrast, the \textbf{taxonomy-based RAG} approach substantially enhances invalid recall and overall consistency. By retrieving semantically related examples with explicit rejection reasons, the model learns to reason from prior invalid patterns instead of relying purely on surface similarity. As shown in Table~\ref{tab:ID_rq2}, this method yields the best overall performance, achieving an accuracy of 0.885 and a Macro-F1 of 0.815.

For \textbf{GPT-5 + Taxonomy RAG + Vendor Scope}, the performance did not continue to improve. 
Instead, a slight drop was observed compared with using taxonomy RAG alone (Macro-F1 = 0.725). 
This suggests that combining multiple forms of external context does not necessarily yield additive benefits, and the interaction between different types of contextual signals may require further investigation.



\noindent\textbf{Case Study of RAG.}
We examine a representative case to show how RAG complements missing contextual information.

Report \#2215434 describes an information disclosure issue where a PoC video unintentionally reveals a hacker’s email address~\cite{hackerone_2215434}.
Because the textual description is brief and the LLM cannot interpret video content, the baseline GPT-5 model incorrectly labeled it as invalid.
With the RAG method, the model retrieved Report \#2134874 as the most similar example, which is the exact report cited inside \#2215434.
The retrieved report provides a detailed description of the same vulnerability type (emails exposed in a PoC video), which compensates for the missing context and helps GPT-5 recognize that \#2215434 indeed demonstrates a valid privacy disclosure.
Table \ref{tab:rag_case} summarizes this case.

\begin{table}[t]
\centering
\caption{Case study: how RAG retrieved a related report that helped correct GPT-5’s judgment.}
\label{tab:rag_case}
\small
\renewcommand{\arraystretch}{1.35}
\begin{tabular}{p{2.4cm}p{5.5cm}}
\toprule
\textbf{Item} & \textbf{Description} \\
\midrule
Target report & \#2215434: PoC video exposes a hacker’s email address in a local folder. \\
Baseline prediction & Misclassified as invalid due to limited textual detail. \\
Retrieved reports & \#2134874: Describes the same exposure pattern (emails visible in PoC video), referenced inside \#2215434. \\
Improvement & The retrieved report supplies explicit textual evidence of the same vulnerability type, clarifying the missing context. \\
Outcome & Model correctly revised decision, classifying \#2215434 as a valid information disclosure. \\
\bottomrule
\end{tabular}
\end{table}

\subsection{Summary of Findings}
Our results show that adding reviewer knowledge can substantially improve the model’s ability to recognize invalid reports.
The taxonomy-based RAG approach provides concrete examples and reasoning cues, which help the model make more consistent judgments and reduce its bias toward accepting weak submissions.
In contrast, vendor scope information alone brings little benefit because it is often too general to guide the classification process.
Overall, structured knowledge that reflects how reviewers actually reason about invalidity proves more useful than static metadata when improving LLM-based bug bounty validation.


\section{RQ3: Does bias exist on reviewers' invalidity judgment? }

\subsection{Motivation}


While the previous experiments mainly focused on textual content, we also explore whether non-content factors might influence reviewers’ final judgments. In practice, report validation is not a purely mechanical process. It often depends on how the report is written, who submitted it, and how reviewers perceive its credibility.
We investigate whether reports with similar content could still receive different outcomes.

\subsection{Methodology and Experiments}
Building on the similarity computation used in RQ2, we identify the top-$k$ semantically closest reports within each weakness category to form comparable pairs. We then remove pairs that share the same status or contain duplicates, retaining only those with a similarity score above 0.8. This filtering ensures that the compared pairs describe essentially the same vulnerability but were evaluated differently during review.

Next, we align each pair’s review outcomes with the corresponding reporters’ reputation scores from HackerOne. 
We assign ordinal values to report statuses according to their impact on reputation \footnote{Following the ranking of HackerOne's closed report states: \textit{Resolved} $>$ \textit{Informative} $>$ \textit{Not Applicable} $>$ \textit{Spam}, which correspond to reputation changes of +7, 0, --5, and --10 respectively.}~\cite{hackerone_report_states},
and treat missing reputation scores as 0.
This enables us to examine whether a “better” outcome (i.e., a higher-priority status) tends to correspond to reporters with a higher reputation.

\begin{figure}[h]
\centering
\includegraphics[width=0.45\textwidth]{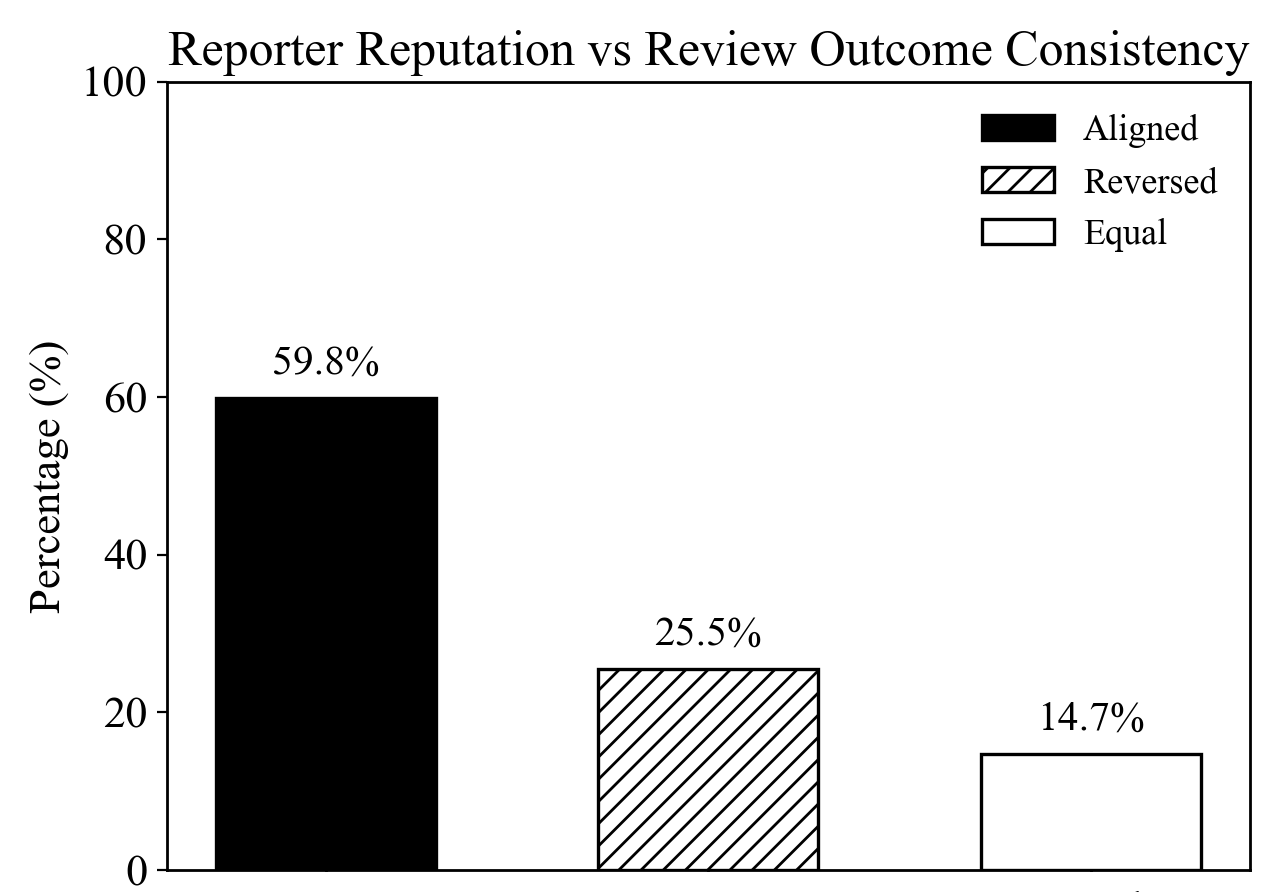}
\caption{\textbf{Aligned}: the higher-reputation reporter received a higher-ranked status (e.g., Resolved over Informative); 
\textbf{Equal}: both had identical reputation; 
\textbf{Reversed}: the higher-reputation reporter received a lower-ranked status.}

\label{fig:reputation_alignment}
\end{figure}

\noindent\textbf{Inconsistency Distribution Analysis}. As shown in Figure~\ref{fig:reputation_alignment}, we analyze a total of 259 \textit{report–neighbor pairs}, each consisting of two semantically similar reports identified within the same weakness category using cosine similarity of their textual embeddings (similarity $\geq 0.8$). Across these pairs, we find that in 155 cases (59.85\%), the report with the \textit{higher status outcome} also came from a reporter with a higher reputation. Among the remaining 104 pairs (40.15\%), 14.67\% cases had identical reputation scores between the two reporters, while 25.48\% cases showed the opposite trend, where the report with a higher-ranked outcome was submitted by a lower-reputation reporter. Although this is not an overwhelming majority, the distribution still shows a clear correlation between a reporter’s reputation and the perceived validity of their submissions. The results suggest that reputation or experience may influence how reviewers interpret and evaluate technically similar reports.

This pattern can also be visually confirmed in Figure~\ref{fig:reputation_alignment}, where the \emph{Aligned} bar dominates the distribution, indicating that reviewers tend to favor experienced reporters when evaluating similar vulnerability reports.


\begin{table*}[htbp]
\centering
\small
\caption{Examples of similar reports with inconsistent reviewer outcomes.}
\label{tab:inconsistency-examples}
\renewcommand{\arraystretch}{1.2}
\begin{tabular}{p{3cm}p{6.7cm}p{6.7cm}}
\toprule
\textbf{Attribute} & \textbf{Case 1} & \textbf{Case 2} \\
\midrule
\multicolumn{3}{l}{\textbf{Example Clickjacking — Different Programs}} \\[2pt]
\textbf{Report ID} & \#688546 (Palo Alto Software)~\cite{hackerone_688546} & \#832593 (Kubernetes)~\cite{hackerone_832593} \\
\textbf{Reporter} & \texttt{paramdham} (reputation = 8,342) & \texttt{hackerboy404} (reputation = 44) \\
\textbf{Target} & \texttt{app.outpost.co} & \texttt{kubernetes.io} \\
\textbf{Vulnerability Type} & Clickjacking (missing \texttt{X-Frame-Options}) & Clickjacking (same issue and PoC) \\
\textbf{Proof of Concept} & Short iframe-based PoC & Similar iframe PoC reused \\
\textbf{Reviewer Decision} & \textit{Resolved} — vendor acknowledged and fixed the issue & \textit{Informative} — no sensitive state change involved \\
\textbf{Outcome} & Accepted and rewarded & Rejected as low impact \\


\bottomrule
\end{tabular}
\end{table*}

\begin{table}[h]
\caption{Fairness evaluation across reputation groups. 
$val|inval$ denotes the percentage of human-accepted reports among model-rejected ones, 
and $inval|val$ denotes the percentage of human-rejected reports among model-accepted ones.
}
\label{tab:boundary_fairness}
\setlength{\tabcolsep}{4pt}

\renewcommand{\arraystretch}{1.3}
\begin{tabular}{lccccl}
\hline
\textbf{Group} & \textbf{TPR} & \textbf{FPR} & \textbf{FNR} & \textbf{$val|inval(\%)$}  &\textbf{$inval|val(\%)$}  \\
\hline
High reputation & 0.965 & 0.694 & 0.035 & 59.26  &5.31\\
Low reputation  & 0.944 & 0.534 & 0.056 &  31.21 &13.01\\
\hline
$\Delta$ (High--Low) & +0.021 & +0.161 & --0.021 & +28.05  &-7.70\\
\hline
\end{tabular}
\end{table}

\begin{figure}[h]
\centering
\includegraphics[width=0.50\textwidth]{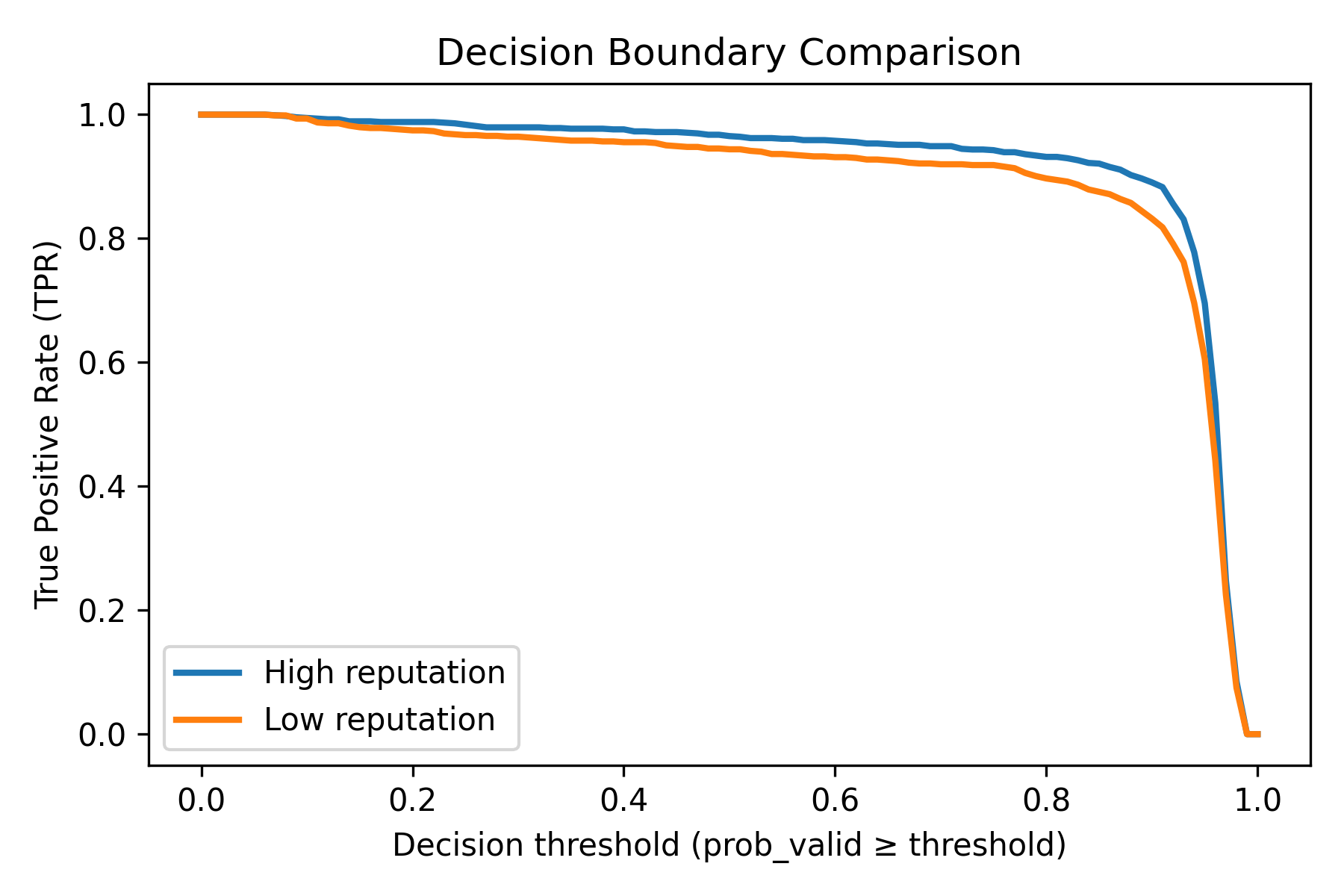}
\caption{Decision boundary comparison between high and low reputation reporters.}
\label{fig:decision_boundary}
\end{figure}

\noindent\textbf{Fairness Test Near the Decision Boundary.}
To evaluate whether reporter reputation influences reviewer leniency, we perform a fairness analysis on \emph{borderline} reports—those that the model predicts as invalid with high confidence but are labeled as valid by human reviewers.

Here we use the fine-tuned RoBERTa model (the best-performing system in RQ1) as a reference to locate these borderline reports. 
We select reports satisfying $P(\text{invalid}) \geq 0.7$ and label = 1 (valid in ground truth), and then divide them into two groups by reporter reputation (high vs. low, split by the median reputation score).

Following prior fairness studies~\cite{hardt2016equality, mitchell2021algorithmic}, we measure potential bias using standard classification metrics:
\begin{itemize}
    \item \textbf{True Positive Rate (TPR)} — proportion of valid reports correctly accepted.
    \item \textbf{False Positive Rate (FPR)} — proportion of invalid-like reports incorrectly accepted.
    \item \textbf{False Negative Rate (FNR)} — proportion of valid-like reports incorrectly rejected.
    \item \textbf{$val \mid inval$} — percentage of human-accepted reports among model-rejected ones:
\[
val \mid inval = 
\frac{\#\{\text{model predicts }0,\ \text{human label }=1\}}
     {\#\{\text{model predicts }0\}}.
\]

\item \textbf{$inval \mid val$} — percentage of human-rejected reports among model-accepted ones:
\[
inval \mid val =
\frac{\#\{\text{model predicts }1,\ \text{human label }=0\}}
     {\#\{\text{model predicts }1\}}.
\]

\end{itemize}

A fair review process should yield similar TPR and FPR across reputation groups.
As shown in Table~\ref{tab:boundary_fairness}, the TPR values for the two groups are close to each other ($\Delta$TPR = 0.02), meaning both groups have a similar rate of correctly accepted valid reports.  
In contrast, the FPR gap is much larger ($\Delta$FPR = 0.16), which suggests that high-reputation reporters are more likely to have invalid-like reports accepted.  
The other two metrics show the same trend: high-reputation reporters have a much higher $val\mid inval$ score (+28.05\%), indicating that model-rejected reports from this group are more often accepted by human reviewers.  
They also have a lower $inval\mid val$ score (–7.70\%), meaning their accepted reports are less likely to be judged invalid by humans.  
Taken together, these results show that reviewers tend to be more lenient toward high-reputation reporters.

This aligns with the intuition that human reviewers apply a more lenient decision boundary to trusted reporters.
The same pattern is reflected in Figure~\ref{fig:decision_boundary}, where the TPR curve for high-reputation reporters consistently lies above that of low-reputation ones.

\noindent\textbf{Example (Clickjacking, Palo Alto Software vs. Kubernetes).}  
From Table~\ref{tab:inconsistency-examples}, we can see that the example compares two reports describing the same clickjacking vulnerability caused by missing \texttt{X-Frame-Options} headers.
Despite nearly identical proof-of-concept code, the results are different: the report submitted by a higher-reputation hunter (\texttt{paramdham}) was accepted and fixed, while the one by a lower-reputation hunter (\texttt{hackerboy404}) was marked as \textit{Informative}.



\subsection{Summary of Findings}
Our analysis provides preliminary evidence that bug bounty reviews may not always be entirely objective.
Reports with nearly identical content can still receive different outcomes, and higher-reputation reporters tend to obtain slightly more favorable results.  
This suggests that reviewer perception, shaped by a reporter’s experience or credibility, may influence decisions in borderline cases.
Understanding this tendency can help future AI-assisted review systems make the process fairer and more consistent.


\section{Related Work}

\subsection{Empirical Studies on Bug Bounty Platforms}
Existing empirical studies have examined bug bounty platforms from various perspectives, including economic efficiency~\cite{zhao2017devising,zhang2025make,walshe2020empirical,allodi2024survey}, participant motivation~\cite{akgul2023bug,barre2025breakers}, and maintainers’ security practices~\cite{ayala2025mixed,ayala2025investigating,ayala2025deep}, while overlooking the review process itself. 

Finifter et al.~\cite{finifter2013empirical} conducted one of the earliest quantitative analyses of vulnerability reward programs, finding that vendor-sponsored bounties were cost-effective compared to hiring additional security staff.  
Akgul et al.~\cite{akgul2023bug} and Barre et al.~\cite{barre2025breakers} investigated hunters’ motivations, highlighting the role of financial incentives, reputation, and learning opportunities, while also identifying demotivating factors such as delayed or unfair evaluations.  
Ayala et al.~\cite{ayala2025deep} examined open-source maintainers on Huntr and observed that reviewers’ acceptance decisions were influenced by report quality and the presence of proof-of-concept (PoC) exploits rather than by formalized criteria.
However, these studies do not analyze how reviewers actually determine validity, what criteria they apply, or how consistent their decisions are.
Our work contributes to filling this gap by analyzing how reviewers make validity decisions and what factors correlate with different outcomes.



In addition, from a methodological standpoint, most prior empirical studies on bug bounty platforms have been conducted at a relatively limited scale.  
For example, Maillart et al.~\cite{maillart2017given} analyzed activity patterns across only 35 public programs on HackerOne, illustrating how vulnerability discoveries decline over time after program launch.  
Similarly, Shafigh et al.~\cite{shafigh2021some} examined just 1,028 fully disclosed invalid reports to develop an out-of-scope taxonomy. 
In contrast, our work utilizes a much larger dataset of 9,942 disclosed reports across 316 programs, which allows for a more comprehensive and statistically grounded analysis of reviewer decisions and invalidity patterns.




\subsection{LLM-Assisted Bug Bounty Analysis}

Research on applying LLMs or agent-based systems to bug bounty review remains scarce.  
Existing work has primarily focused on using LLMs to support hunters in vulnerability discovery and exploit automation.  
For instance, \textit{The Hacker’s Guide to LLMs}~\cite{mohammad2024hacker} explores how LLMs can be used for attack simulation, prompt-based exploitation, and adversarial testing within bug bounty workflows.
However, these studies address the offensive and discovery aspects of bug bounty processes rather than the review and validation stage.  

Our work investigates the use of LLMs as review agents that leverage natural language reasoning and reviewer knowledge (such as report taxonomies and program scope) to analyze report validity.  
To the best of our knowledge, no prior empirical study has systematically modeled human reviewer decisions using LLMs or agent-based reasoning in the context of bug bounty report validation.

\section{Conclusion}

This paper studied how bug bounty reviewers decide the validity of vulnerability reports and explored whether LLMs can assist in this process.  
After collecting 9,942 disclosed HackerOne reports, we evaluated GPT-5, DeepSeek, and a fine-tuned RoBERTa model on report validity prediction.  
Although the models performed well on valid reports, they often failed to recognize invalid ones, showing a clear tendency toward acceptance.  
Adding reviewer knowledge through a taxonomy-based retrieval method improved both recall and consistency, suggesting that structured contextual cues help models make more balanced judgments.  
Our analysis of similar reports with different outcomes also indicates that reviewer perception and reporter reputation may influence validation results.

\section{Limitations and Future Work}
Although our study provides new empirical insights, it has several limitations.

First, our current taxonomy focuses only on the \textit{Information Disclosure} category, which limits the generality of our conclusions.  
We plan to expand it to cover other major vulnerability types, including cross-site scripting, access control, and authentication issues, to build a more comprehensive reviewer knowledge base. 

Furthermore, our dataset includes only publicly disclosed reports on HackerOne; private programs may exhibit different review dynamics.
We will extend our dataset to include reports from other bug bounty platforms such as Bugcrowd and Huntr, as well as private disclosure programs.  
A larger and more diverse dataset will allow for a broader analysis of reviewer behavior and improve model generalization across platforms.  

In addition, our evaluation relies on text-based similarity and may overlook non-textual cues such as proof-of-concept videos or attachments. These aspects will be addressed in our future work.

Through these efforts, we aim to better understand how reviewer knowledge and cross-platform data can enhance the fairness and consistency of bug bounty validation.

\section{Ethical Considerations}

All data used in this study come from publicly disclosed HackerOne reports. 
We analyze only information that is voluntarily published by reporters and vendors, and we do not attempt to identify individuals or make judgments about specific reviewers. 
All results are presented in aggregate, and examples are used solely for scientific illustration. 
Because no private data or human intervention was involved, this work is exempt from institutional review board (IRB) oversight under standard guidelines.

\clearpage
\bibliographystyle{IEEEtran}
\bibliography{IEEEabrv}

\end{document}